\documentclass[epj]{webofc}
\usepackage[varg]{txfonts}   % Web of Conferences font
\wocname{EPJ Web of Conferences}
\woctitle{INPC 2013}

\begin{document}

\title{The nucleon mass and pion-nucleon sigma term from a chiral analysis of $N_f = 2$ lattice QCD world data}

\author{L. Alvarez-Ruso\inst{1} \and
        T. Ledwig\inst{2}\fnsep\thanks{\email{ledwig@ific.uv.es}} 
        J. Martin Camalich\inst{3} \and
        M. J. Vicente Vacas\inst{2}
}

\institute{
  Instituto de F\'isica Corpuscular (IFIC), Centro Mixto
  Universidad de Valencia-CSIC, Valencia, Spain 
  \and
  Departamento de F\'\i sica Te\'orica and IFIC, Centro Mixto
  Universidad de Valencia-CSIC, Valencia, Spain
  \and
  Department of Physics and Astronomy, University of Sussex, Brighton, UK
}

\abstract{
  We investigate the pion-mass dependence of the nucleon mass within the covariant $SU(2)$ baryon chiral
  perturbation theory up to order $p^4$ with and without explicit $\Delta\left(1232\right)$ degrees of 
  freedom. We fit lattice QCD data from several collaborations for 2 and 2+1 flavor 
  ensembles. Here, we emphasize our $N_f=2$ study where the inclusion
  the $\Delta(1232)$ contributions stabilizes the fits. We correct for finite volume and 
  spacing effects, set independently the lattice QCD scale by
  a Sommer-scale of $r_0=0.493(23)$ fm and also include one $\sigma_{\pi N}$ lQCD data point at 
  $M_\pi \approx 290$ MeV. We obtain low-energy constants of natural size which are compatible with 
  the rather linear pion-mass dependence observed in lattice QCD.
  We report a value of 
  $\sigma_{\pi N}=41(5)(4)$~MeV for the 2 flavor case and $\sigma_{\pi N}=52(3)(8)$ MeV for 2+1 flavors.
}

\maketitle

\section{Introduction}\label{intro}
The current quark mass dependence of observables from lattice QCD simulations (lQCD) 
for unphysical values provides an additional perspective on QCD itself.
Many simulations with two dynamical degenerated light quarks ($N_f=2$) 
or two degenerated light and one heavy quark ($N_f=2+1$) are now available
with light quark masses spanning from unphysical heavy values down to nearly physical ones. 
To investigate this extensive data set the (baryon) chiral perturbation theory (B$\chi$PT)
represents a prominent tool. By matching B$\chi$PT to lQCD data for unphysical quark masses, 
low-energy constants (LECs) can be extracted, which are then used for predictions 
at the physical point.

We performed such a matching for the quark-mass ($m_{u}=m_{d}=\overline{m}$) dependence of the nucleon 
mass $M_N(\overline{m})$ by 
separately fitting to $N_f=2$ and $2+1$ lQCD data. We concentrate here on our 
$N_f=2$ results and refer to \cite{paper} for more details on the $2+1$ ones.

An important derivative of $M_N(\overline{m})$ is given by the Hellmann-Feynman theorem:
\begin{equation}
  \overline{m}\frac{\partial}{\partial\overline{m}}M_{N}\left(\overline{m}\right)=\sigma_{\pi N}=
  \overline{m}\langle N |\overline{u}u+\overline{d}d| N \rangle\,\,\,\,,
\label{eq:HF}
\end{equation}
which relates $M_N(\overline{m})$ to the so-called $\sigma_{\pi N}$ term. The $\sigma_{\pi N}$ term 
can be seen as a measure 
of the contribution from the explicit chiral symmetry breaking to the nucleon mass. Since the 
$\sigma_{\pi N}$ can also be defined by the nucleon scalar form factor at zero four-momentum 
transfer squared, it can also be isolated from $\pi N$-scattering data. 
In this sense, the quark-mass dependence of the 
nucleon mass relates lQCD, B$\chi$PT and experiment.

\section{Nucleon mass and covariant baryon chiral perturbation theory} \label{ChPT}

The chiral structure of the nucleon mass is parametrized by the covariant B$\chi$PT up to order $p^4$ as:
\begin{equation}
  M_{N}^{\left(4\right)}\left(M_{\pi}^{2}\right)  =  M_{0}-c_{1}4M_{\pi}^{2}+\frac{1}{2}
  \overline{\alpha}M_{\pi}^{4}+\frac{c_{1}}{8\pi^{2}f_{\pi}^{2}}M_{\pi}^{4}
  \ln\frac{M_{\pi}^{2}}{M_{0}^{2}}+\Sigma_{loops}^{\left(3\right)+\left(4\right)}\left(M_{\pi}^{2}\right)
  +\mathcal{O}\left(p^{5}\right)\,\,.\label{eq:MN(4)(Mpi)}
\end{equation}
where $M_\pi^2 \sim \overline{m}$ and $f_\pi$ are the pion mass and pion decay constant. The loop-contributions 
$\Sigma_{loops}^{\left(3\right)+\left(4\right)}$ can
also contain explicit $\Delta(1232)$ contributions. We refer to Ref. \cite{paper} for all the details. 
We fit the nucleon chiral-limit mass $M_0$ and the two LECs $c_1$ and $\overline{\alpha}$ to 
lQCD data and obtain through Eq. (\ref{eq:HF}) a 
$\sigma_{\pi N}$ value. Explicitly, we use the $N_f=2$ lQCD data in its dimensionless 
from $\left(r_0 M_\pi, r_0 M_N\right)$, with $r_0$ being the Sommer-scale, and minimize the function: 
\begin{eqnarray}
  \chi^{2} & = & \sum_{i}\left[\frac{\widetilde{M}_{N}^{(4)}\left(\widetilde{M}_{\pi}^{2}\right)+
      \widetilde{\Sigma}^{(4)}_{N}\left(\widetilde{M}_{\pi}^{2},L \right)+\tilde{c}_{a}\tilde{a}^{2}-d_{i}
      \left(\widetilde{M}_{\pi}^{2},L \right)}{\sigma_{i}}\right]^{2}\,\,\,\,,\label{eq:chi2 nf2}\\
  \mbox{with} &  & \widetilde{M}_{N}^{(4)}=r_{0}M_{N}^{(4)}\,\,\,\,,\,\,\,\,\widetilde{M}_{\pi}^{2}=
  \left(r_{0}M_{\pi}\right)^{2}\,\,\,\,,\,\,\,\,\widetilde{\Sigma}_{N}^{(4)}=r_{0}\Sigma_{N}^{(4)}\,\,\,\,, 
\end{eqnarray}
where $d_{i}\left(\widetilde{M}_{\pi}^{2},L\right)$ are the data points with uncertainties $\sigma_{i}$ 
for a lattice of size $L$ and spacing $a$. The terms $\tilde{c}_a a^2$ parametrize finite spacing effects
for each lQCD action separately. The self-energy $\Sigma_{N}^{(4)}$ contains 
also finite volume corrections. Furthermore, we use the physical nucleon mass to determine
the Sommer-scale $r_0$ recursively and self-consistently inside the fit.
In our $SU(2)$ fits to $N_f=2+1$ data we assume that the strange quark contributions are integrated out
and absorbed into the LECs.

To ensure controlled finite volume effects and an acceptable chiral convergence of our B$\chi$PT results, 
we restrict our fits to $M_N$-data points 
fulfilling $M_{\pi} L > 3.8$ and  $r_{0} M_{\pi} < 1.11$. Additionally, the QCDSF collaboration
obtained one direct $\sigma_{\pi N}$ data point at $M_\pi\approx 290$ MeV \cite{QCDSF} 
with which we also perform simultaneous fits to nucleon mass data and that $\sigma_{\pi N}(290)$ point.

\section{Results and conclusions}\label{Results}

\begin{figure}[h]
\centering
\includegraphics[width=7cm,clip]{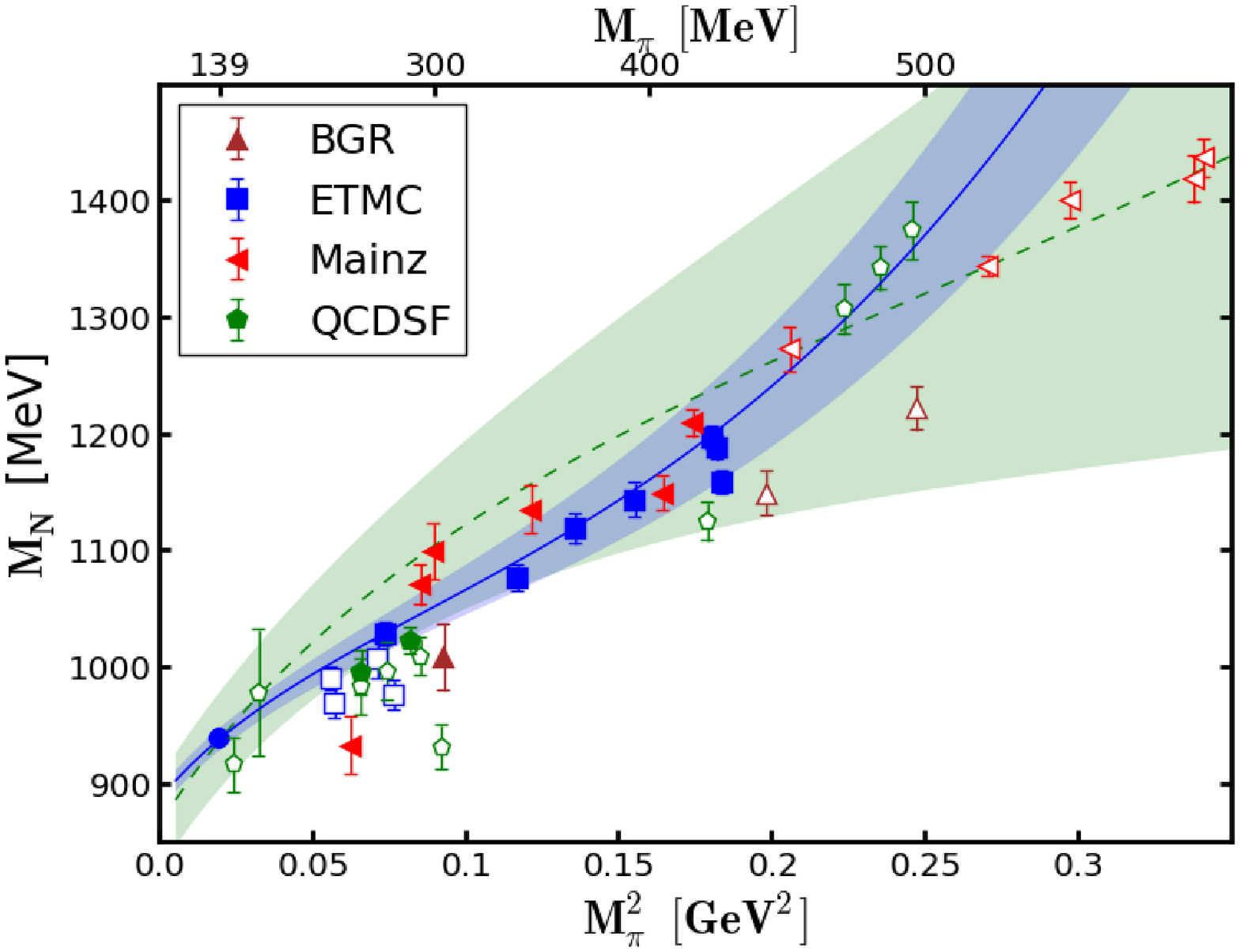}\includegraphics[width=7cm,clip]{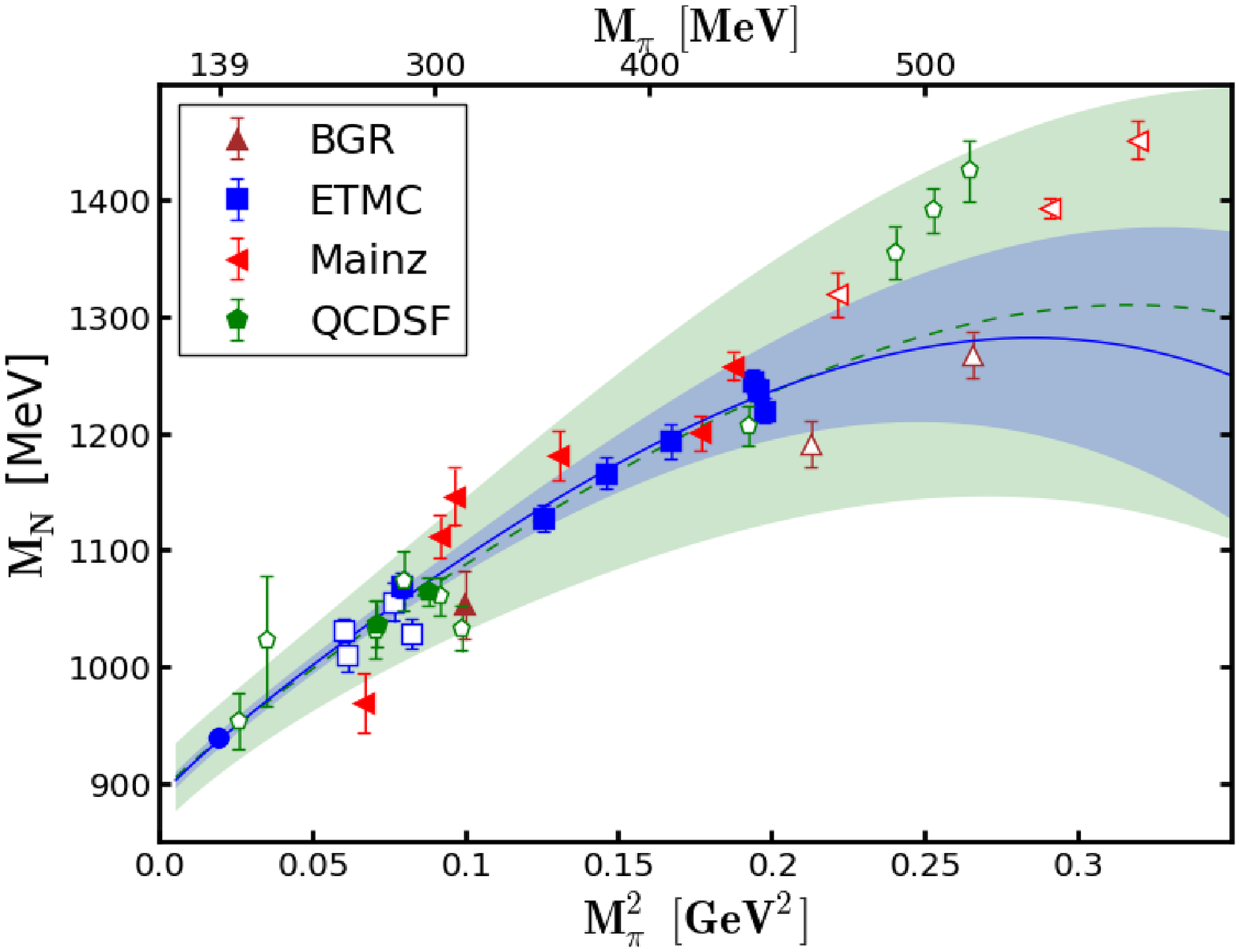}
\caption{Pion-mass dependence of the nucleon mass from $N_f=2$ fits. We show B$\chi$PT $p^4$ fits without (left)
  and with (right) explicit $\Delta(1223)$ degrees of freedom. 
  The green-dotted lines correspond to fits to nucleon mass data alone and the blue-solid lines 
  to simultaneous fits including also the $\sigma_{\pi N}(290)$ QCDSF data point. The lQCD data points are 
  scaled by our $r_0 \approx 0.490$ fm obtained in the fits with $\Delta(1232)$.}
\label{fig-1}     
\end{figure}

We summarize some of our results of Ref. \cite{paper}.
Figure \ref{fig-1} shows our fits to the lQCD data from the $N_f=2$ ensembles of the 
BGR, ETMC, Mainz and QCDSF 
collaborations \cite{BGR, ETMC, Mainz, QCDSF}. 
We fitted the data without and with explicit $\Delta(1232)$ contributions, 
left and right figures respectively. The dashed and solid lines correspond to the ex-/inclusion of the 
one $\sigma_{\pi N}(290)$ point, respectively. Our fits yield $2<\chi^2/d.o.f.<3$ reflecting that some 
of the data are marginally consistent.

By including the one $\sigma_{\pi N}(290)$ point in the $\Delta(1232)$-less fit we reduce the uncertainties, 
 although, the shape of the pion-mass dependence changes noticeably. This is also seen
in the obtained $\sigma_{\pi N}$ value at the physical point, which changes from $\sigma_{\pi N}=62(13)$ MeV 
to $41(3)$ MeV. The situation is different when we include the explicit $\Delta(1232)$ contributions in 
our fit formula. For this case, the inclusion of the one $\sigma_{\pi N}(290)$ point does not change the pion-mass 
dependence much. Differences are mainly visible at higher pion masses and the $\sigma_{\pi N}$ term at the 
physical point turns out to be $\sigma_{\pi N}=41(3)$ MeV for both cases. We conclude that for the present
data situation the inclusion of $\Delta(1232)$ contributions stabilizes the fits and that the reported 
(only one) $\sigma_{\pi N}(290)$ point is more compatible with the B$\chi$PT with $\Delta(1232)$ rather 
than with the $\Delta(1232)$-less one. It will be interesting so see if further direct lQCD calculations 
of the $\sigma_{\pi N}$ at unphysical pion-masses will confirm this conclusion.

\begin{figure}[h]
\centering
\includegraphics[width=7cm,clip]{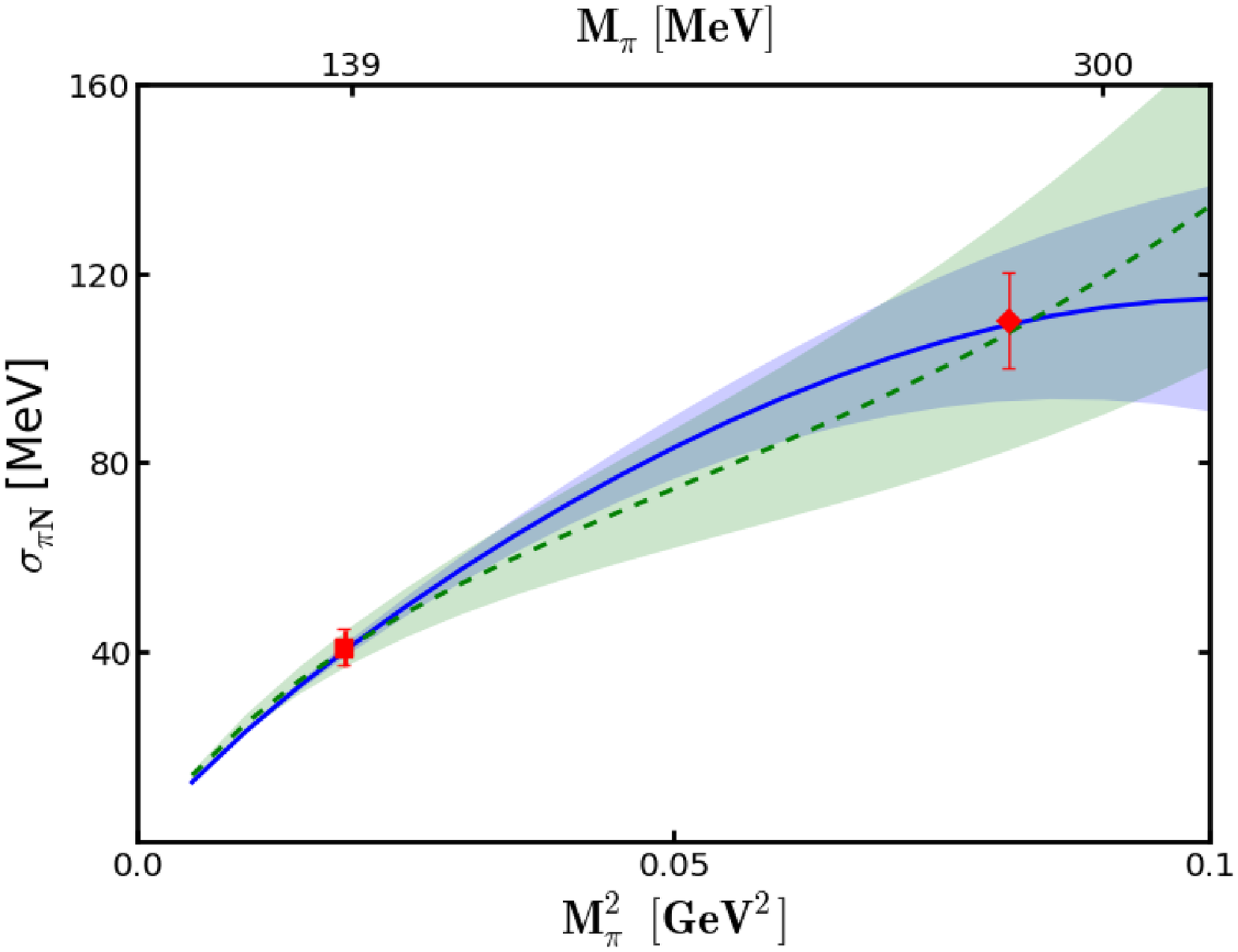}\includegraphics[width=7cm,clip]{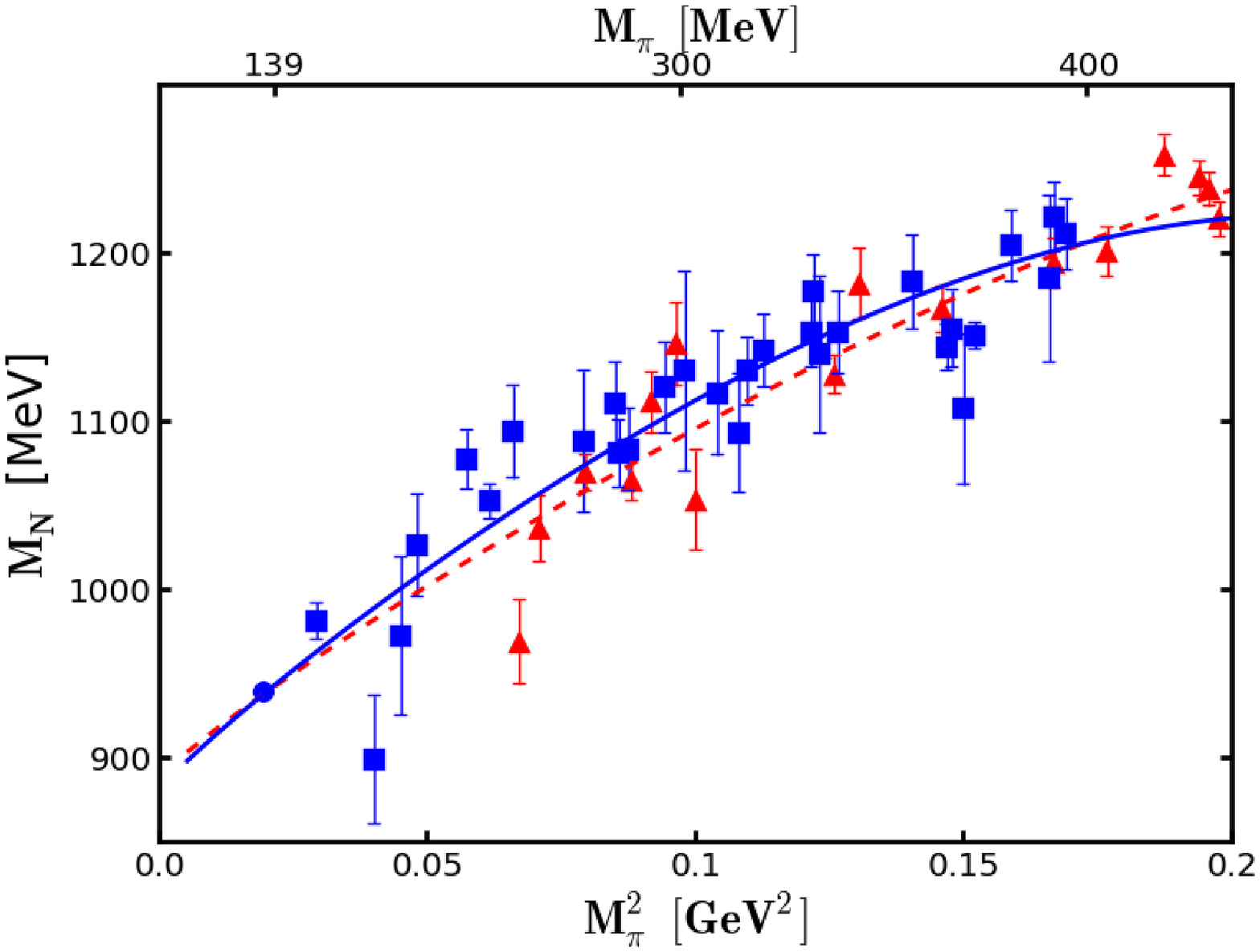}
\caption{Left: Pion-mass dependence of the $\sigma_{\pi N}$ term from $N_f=2$ fits. The blue-solid (green-dotted) line 
  corresponds to the  B$\chi$PT $p^4$ fit with(out) the $\Delta(1232)$  with including the 
  $\sigma_{\pi N}(290)$ data point (red-diamond). The red-square is our predicted $\sigma_{\pi N}$ value 
  at the physical point. Right: Comparison of B$\chi$PT $p^4$ fits with $\Delta(1232)$ to $N_f=2$ 
  (red-triangles) and $N_f=2+1$ (blue-squares) lQCD data. The blue-circle is the physical nucleon mass.}
\label{fig-2}     
\end{figure}

The left panel of Fig. \ref{fig-2} shows the pion-mass dependence of the $\sigma_{\pi N}$ term 
obtained from our B$\chi$PT fits with $\Delta(1232)$ contributions, solid line, and without, dashed-line.
 As discussed above, the changes of the slope of the dashed line could indicate that the 
$\Delta(1232)$-less B$\chi$PT has problems to account for the $\sigma_{\pi N}(290)$ point. 
As for our final $\sigma_{\pi N}$ value at the physical point for the $N_f=2$ fits we quote:
\begin{equation}
  \sigma_{\pi N}^{N_f=2}=41(5)(4)\,\,\textrm{MeV}\,\,, 
\end{equation}
which corresponds to our result for the B$\chi$PT with $\Delta(1232)$ contributions. 
For the determination of the uncertainties we refer to Ref. \cite{paper}.

The right panel of Fig. \ref{fig-2} compares our B$\chi$PT results from fits to $N_f=2$ and
 $N_f=2+1$ data. Differences are within
the errorbars of the input data but translate into a $\sim 11$ MeV 
difference in the $\sigma_{\pi N}$ value. We obtain for our $N_f=2+1$ fits the higher value of
\begin{equation}
  \sigma_{\pi N}^{N_f=2+1}=52(3)(8)\,\,\textrm{MeV}\,\,.
\end{equation}
Both values are compatible within the uncertainties, however, with the present
data we cannot unambiguously determine the origin of the $11$ MeV 
difference of the central values. 
As a first point, we see that the $N_f=2$ data does not constrain the small pion-mass 
region much. This results in an up to $\sim13$ \% smaller $c_1$ value which determines solely 
the $\sigma_{\pi N}$ at leading order. As a second point, the $\sigma_{\pi N}(290)$ data point brought in the 
$N_f=2$ case the $\Delta(1232)$ and $\Delta(1232)$-less results together. Such low-$M_\pi$ $\sigma_{\pi N}$ 
data points are not available
for the $N_f=2+1$ case and our fits yield by $9 \sim 11$ MeV different $\sigma_{\pi N}$
 results for the two cases. The above $N_f=2+1$ value is the average of our $\Delta(1232)$ and 
$\Delta(1232)$-less results. Furthermore, since the strange quark is treated differently in
the two lQCD data sets, one could expect that part of the difference comes also from this fact.
Note also that the latest value extracted from pure $\pi$-N scattering data yields
$\sigma_{\pi N} = 59(7)$ MeV \cite{piN}.

In summary, we fitted lQCD data for $N_f=2$ and $N_f=2+1$ ensembles by a $SU(2)$ B$\chi$PT formula up to
$p^4$ with and without $\Delta(1232)$ degrees of freedom. Even though the present data set
is extensive, we observed systematic uncertainties stemming mostly from the distribution of the data points.
New data for the following cases would further reduce these systematic effects: 
a) more $N_f=2$ data points in the low-$M_\pi$ region, 
b) even one direct calculation of the $\sigma_{\pi N}$ at $M_\pi<300$ MeV for the $N_f=2+1$ case.

\section*{Acknowledgements}
The work has been supported by the Spanish Ministerio de Economía y Competitividad and 
European FEDER funds under Contracts FIS2011-28853-C02-01 and FIS2011-28853-C02-02, Generalitat Valenciana 
under contract PROMETEO/2009/0090 and the EU Hadron-Physics3 project, Grant No. 283286. JMC also 
acknowledges support from the Science Technology and Facilities Council (STFC) under grant ST/J000477/1,
the grants FPA2010-17806 and Fundaci\'on S\'eneca 11871/PI/09.

\end{document}